\documentclass[aps,prl,epsf,epsfig,twocolumn]{revtex4}

\usepackage[dvips]{graphicx}

\begin{document}

%@twocolumnfalse\endcsname

\title{Kohn-Sham Exchange Potential for a Metallic Surface}
\author{C. M. Horowitz, C. R. Proetto and S. Rigamonti}
\address{Centro At\'{o}mico Bariloche and Instituto Balseiro, 8400 S. C. de
Bariloche, R\'{\i}o Negro, Argentina}

\begin{abstract}
The behavior of the surface barrier that forms at the metal-vacuum interface 
is important for several fields of surface science. Within the Density 
Functional Theory framework, this surface barrier has two non-trivial components: 
exchange and correlation. Exact results are provided for the exchange 
component, for a jellium metal-vacuum interface, in a slab geometry.     
 The Kohn-Sham exact-exchange potential $V_{x}(z)$ has been 
generated by using the Optimized Effective Potential method, through an 
accurate numerical solution, imposing the
correct boundary condition. It has been proved analytically, and confirmed numerically,
that $V_{x}(z\rightarrow \infty )\rightarrow -\,e^{2}/z$; this conclusion is not affected by the inclusion of correlation effects. Also, the 
exact-exchange potential develops a shoulder-like structure close to the interface, on the
vacuum side. The issue of the classical image potential is discussed.   
\end{abstract}

\maketitle

%\vskip

Density Functional Theory (DFT) is a successful theory to calculate 
the electronic structure of atoms, molecules, clusters, and solids. Its goal is 
the quantitative understanding of materials properties starting from the fundamental 
laws of quantum mechanics. It is then a
major drawback of DFT\cite{parr}, that when
applied in its highly successful Local Density Approximation (LDA) to the 
metal-vacuum interface system, it yields an exponential vanishing exchange-correlation (xc) potential which fails to
reproduce the image-like asymptotic behavior of the surface barrier\cite{lang}. 
This problem of LDA is common to all local or semi-local
extensions of it (GGA, meta-GGA,...). More importantly, the issue of the
long-range behavior of the surface barrier is not even settled from the
conceptual point of view, being still unclear the relative importance of
exchange and correlation in determining this image-like decay\cite{dobson}.
The aim of this work is to provide a rigorous state-of-the-art calculation of
the exchange component of the Kohn-Sham surface barrier for the simplest
model of a jellium metal-vacuum interface. We have found that $V_{x}(z)$
behaves as $-\,e^{2}/z$ for large $z$ in the vacuum region, and that it
presents a shoulder close to the interface, although mainly located
in the vacuum side. 
These findings are of great importance for the interpretation 
of a variety of surface sensitive experiments\cite{dobson}.

Our calculations are restricted to the slab-jellium model of a metallic
surface, where the discrete character of the positive ions inside the metal
is replaced by a uniform distribution of positive charge (the jellium). The
positive jellium density is given by $n_{+}(z)=\overline{n}\,\theta
(d/2-\left| z+d/2\right| )$, which describes a slab of width $d,$ with
jellium edges at $z=-d,0.$ The model is invariant under translations in the $%
x,y$ plane (area $A$), so the wave-functions of the auxiliary Kohn-Sham
system can be factorized as $\varphi _{i{\bf k}}({\bf r})=e^{i{\bf k\cdot
\rho }}\xi _{i}(z)/\sqrt{A}$, where ${\bf \rho }$ and ${\bf k}$ are the
in-plane coordinate and wave-vector, respectively. $\xi _{i}(z)$ are the
normalized spin-degenerate eigenfunctions for electrons in slab discrete
levels $i$ $(=1,2,...),$ and energy $\varepsilon _{i}.$ Within the Kohn-Sham
implementation of DFT, they are the solutions of 
\begin{equation}
\widehat{h}_{\text{KS}}^{i}(z)\xi _{i}(z)\equiv \left[ \frac{-\hbar ^{2}}{%
2m_{0}}\frac{\partial^{2} }{\partial z^{2}}+V_{\text{KS}}(z)-\varepsilon
_{i}\right] \xi _{i}(z)=0.  \label{KS}
\end{equation}
The KS potential is the sum of several contributions: $V_{\text{KS}}(z)=V_{%
\text{H}}(z)+V_{xc}(z).$ $V_{\text{H}}(z)$ is the classical (electrostatic)
Hartree potential. $V_{xc}(z)$ is the non-classical $xc$ contribution; it is
given by $V_{xc}(z)=\delta E_{xc}/\delta n(z).$ $E_{xc}\equiv $ $%
E_{xc}\left[ \left\{ \varepsilon _{i}\right\} \left\{ \xi _{i}\right\}
\right] $ is the $xc$ contribution to the total energy-functional, and $%
n(z)=\sum_{i}^{occ.}(k_{F}^{i})^{2}\left| \xi _{i}(z)\right| ^{2}/2\pi $ is
the 3D density, with $k_{F}^{i}=\sqrt{2m_{0}(\varepsilon _{F}-\varepsilon
_{i})}/\hbar .$ $\varepsilon _{F}$ is the metal Fermi energy, given by the
neutrality condition $\varepsilon _{F}=\hbar ^{2}k_{F}^{2}/2m_{0}+
V_{\text{KS}}(-d/2)$, with $%
k_{F}=(3\pi ^{2}\overline{n})^{1/3}.$ The Optimized Effective Potential
(OEP) method of DFT has been specially designed for dealing with
wave-function and eigenvalue dependent $E_{xc}$, as is our case\cite{kli}.
After some lengthy but standard manipulations of the OEP scheme, the
calculation of $V_{xc}(z)=$ $V_{xc,1}(z)+V_{xc,2}(z)$ for real $\xi _{i}(z)$'s 
and $E_{xc}$ functionals which only depends on occupied subbands 
can be summarized in the following set of equations\cite{epl}, 
\begin{eqnarray}
V_{xc,1}(z) &=&\sum_{i}^{occ.}\frac{\left[ k_{F}^{i}\xi _{i}(z)\right] ^{2}}{%
2\pi n(z)}\left[ u_{xc}^{i}(z)+\Delta \overline{V}_{xc}^{i}\right] ,
\label{xc1} \\
V_{xc,2}(z) &\!\!\!\!=&\!\!\!\!\sum_{i}^{occ.}(\varepsilon _{F}\!\!-\!\!
\varepsilon _{i})\frac{%
(k_{F}^{i})^{2}\psi _{i}(z)\xi _{i}(z)\!+\!\psi _{i}^{\prime }(z)\xi
_{i}^{\prime }(z)}{\pi n(z)},  \label{xc2}
\end{eqnarray}
where the ``shifts'' $\psi _{i}(z)$ are given by 
\begin{equation}
\psi _{i}(z)=\sum_{j\neq i}\xi _{j}(z)\int_{-\infty }^{\infty }\xi
_{j}(z^{\prime })\frac{\Delta V_{xc}^{i}(z^{\prime })}{(\varepsilon
_{j}-\varepsilon _{i})}\xi _{i}(z^{\prime })dz^{\prime },  \label{shift}
\end{equation}
with primes denoting derivatives with respect to $z.$ Here, $\Delta
V_{xc}^{i}(z)=V_{xc}(z)-u_{xc}^{i}(z),$ $u_{xc}^{i}(z)=[4\pi
/A(k_{F}^{i})^{2}\xi _{i}(z)]\delta E_{xc}/\delta \xi _{i}(z),$ and mean
values (for later use) are defined as $\overline{O}^{i}=\int \xi
_{i}(z)O(z)\xi _{i}(z)dz.$ Eqs.(\ref{KS})-(\ref{xc2}) have
to be solved self-consistently. Several comments are worth here: {\it a)}
Eqs.(\ref{xc1})-(\ref{shift}) are a set of integral equations for the local
(multiplicative) $xc$ potential; 
{\it b)} The shifts are invariant under the replacement $
V_{xc}(z)\rightarrow V_{xc}(z)+\alpha ,$ with $\alpha $ a constant. This
means that the above set of equations determines $V_{xc}(z)$ up to an
additive constant, that should be fixed by imposing some suitable boundary
condition. This is a general property of all DFT calculations for fixed
number of particles, as is the present case; {\it c)} If the shifts are forced 
to be identically zero, the only term
that survives is $V_{xc,1}(z).$ This is exactly the KLI
approximation\cite{kli}, which brings the identification $%
V_{xc,1}(z)=V_{xc}^{\text{KLI}}(z).$ All results given until this point
include both exchange and correlation. Unless stated otherwise, we will
concentrate now in the $x$-only case, where $E_{xc}\rightarrow E_{x}$\cite{reboredo}.

The long-range behavior of $V_{x}(z)$ in the vacuum region is an important
point, that could be obtained for our slab geometry 
directly from Eqs.(\ref{xc1}), (\ref{xc2}). For
this, first note that by assuming that $V_{\text{KS}}(z\rightarrow \infty
)\rightarrow 0$ (which is equivalent to the assumption that $%
V_{x}(z\rightarrow \infty )\rightarrow 0$), from Eq.(\ref{KS}) we obtain
that $\xi _{i}(z\rightarrow \infty )\rightarrow e^{-\,z\sqrt{%
-2m_{0}\varepsilon _{i}}/\hbar }$ for all occupied $i$ (disregarding a 
factor involving powers of 
$z$). Following the analysis of Refs.\cite{kriebich} and\cite{gorling}, one
can derive also that $\psi _{i<m}(z\rightarrow \infty )\rightarrow e^{-\,z\sqrt{%
-2m_{0}\varepsilon _{m}}/\hbar },$ and that $\psi _{m}(z\rightarrow \infty
)\rightarrow e^{-\,\,z\sqrt{-2m_{0}\varepsilon _{m-1}}/\hbar }.$ Here, $i=m$
is the last occupied slab discrete level.  Armed with these results, the
asymptotic limit of $V_{x}(z)$ is immediate: $V_{x,2}(z)$ tends
exponentially to zero, while $V_{x,1}(z\rightarrow \infty )\rightarrow
u_{x}^{m}(z)+\Delta \overline{V}_{x}^{m}$. Besides, as $u_{x}^{m}(z%
\rightarrow \infty )\rightarrow 0$ (see below), we conclude that $%
V_{x}(z\rightarrow \infty )\rightarrow V_{x,1}(z\rightarrow \infty
)\rightarrow \Delta \overline{V}_{x}^{m}.$ Consistency with the starting
assumption $V_{x}(z\rightarrow \infty )\rightarrow 0,$ yields the important
constraint $\Delta \overline{V}_{x}^{m}=\overline{V}_{x}^{m}-\overline{u}%
_{x}^{m}=0.$ This constraint fixes the undetermined constant in $V_{x}(z)$  
discussed above. All numerical results of this work have
been obtained by using this constraint as boundary condition. We have
achieved the self-consistent numerical solution of Eqs.(\ref{KS})-(\ref{xc2}) 
by two different methods: $i)$ direct calculation of the
shifts\cite{kummel}, through the solution of the inhomogeneous differential
equation which results from application of the operator $\widehat{h}_{\text{%
KS}}^{i}(z)$ to the shifts of Eq.(\ref{shift}); and $ii)$ direct solution of
the OEP integral equation for $V_{x}(z),$ that is exactly equivalent and can
be obtained from Eqs.(\ref{xc1})-(\ref{xc2})\cite{krotscheck}. Both methods
agree in their results within numerical accuracy, although the first
approach using the shifts is faster in computer time than the second. Both
methods face numerical instabilities 
beyond a critical coordinate $z$ in the vacuum region.

\begin{figure}[h]
\includegraphics[width=\linewidth,clip=true,angle=0]{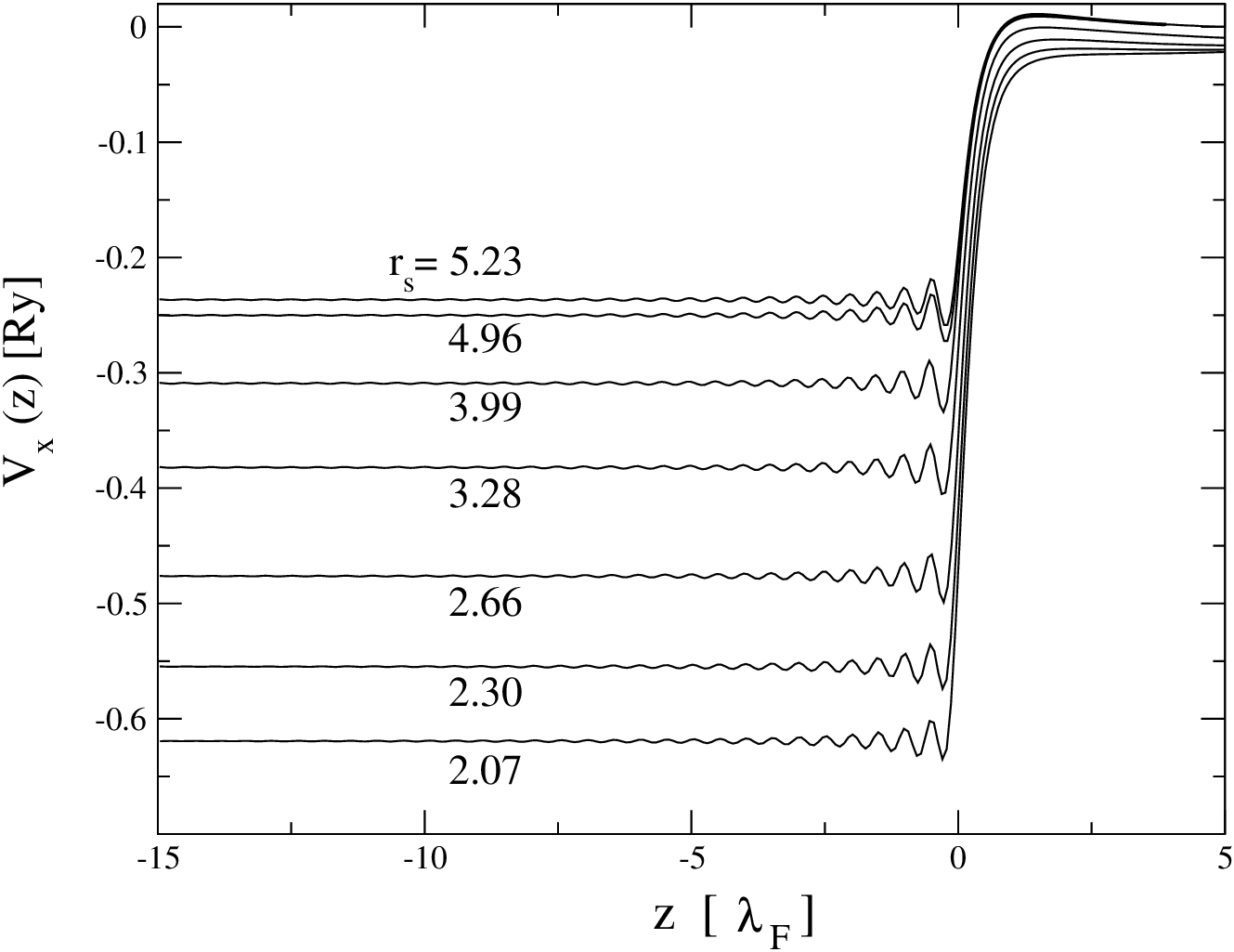}
\caption{Dependence of $V_{x}(z)$ on jellium density, for a fixed slab width 
$d=30$ $\lambda _{F}.$ Jellium edge at $z=0,$ slab center at $z=-d/2=-15$ $%
\lambda _{F}$. $z\geq 0$ corresponds to the vacuum region. Note that as $%
\lambda _{F}=(32\pi ^{2}/9)^{1/3}r_{s}a_{0},$ the thickness of each slab (in
units of $a_{0}$) increases from bottom to top.}
\end{figure}

We start by presenting in Fig.1 numerical results for $V_{x}(z)$, running
from relatively high (Al) to low (Rb) typical metallic densities\cite{units}.
The exact-exchange potential shows large-amplitude oscillations in the
metallic side close to the jellium edge\cite{lang}, strongly depends on
density in the bulk-like region at the slab center, and develops a 
``shoulder'' close to the jellium edge, on the vacuum side. For an homogeneous 3D
electron gas $V_{x}(z)$ becomes a constant, given by $V_{x}($3D$)=-(18/\pi
^{2})^{1/3}/r_{s}\simeq -0.122/r_{s}.$ Replacing in this expression for $%
V_{x}($3D$)$ the $r_{s}$ values of Fig.1, we obtain $-0.590$ (Al), $-0.531$
(Pb), $-0.459$ (Mg), $-0.372$ (Li), $-0.306$ (Na), $-0.246$ (K), and $-0.234$
(Rb). The results of Fig.1 for $V_{x}(-d/2)$ are close to these numbers,
although  they are systematically more negative, due to a slab
finite-size effect. 

Two striking features of $V_{x}(z)$ remain to be discussed: $i)$
the building of a shoulder structure close to the metal-vacuum interface,
and $ii)$ the long-range behavior far from the jellium edge. The strength of
the shoulder structure depends on density (Fig.1) and slab size 
(see top panel of Fig.3). We show in Fig.2 the
details of the shoulder structure: it is due to the shift-dependent term in $%
V_{x}(z)$, that is, $V_{x,2}(z).$ This contribution is very small in the
bulk-like region at the slab center, but exhibits oscillations when
approaching the jellium edge, yielding the shoulder in the total
exact-exchange potential right after the interface. It is important to note
that this effect is beyond the KLI approximation, which amounts to
approximate $V_{x}(z)$ by $V_{x,1}(z).$

The detailed asymptotic behavior of the exact-exchange potential is best discussed
starting from the previous result that $V_{x}(z\rightarrow \infty
)\rightarrow u_{x}^{m}(z\rightarrow \infty ).$ Using the exact-exchange
energy functional appropriate for a slab geometry\cite{reboredo}, we obtain 

\begin{equation}
u_{x}^{m}(z\rightarrow \infty )\rightarrow
-e^{2}k_{F}^{m}\int\limits_{-\infty }^{\infty }\xi _{m}^{2}(y)\,
F(k_{F}^{m}\left|z-y\right|)\,dy,  \label{HF1}
\end{equation}
with $F(x)=\left[x+L_{1}(2x)-I_{1}(2x)\right]/x^{2}$, and $I_{1}$ and $L_{1}$ being the modified Bessel and Struve
functions, respectively. Considering now that in the asymptotic limit 
$z \gg y$, an expansion of the functions $I_{1}$ and 
$L_{1}$ in the limit of large arguments leads to 
$ F(z\gg y) \rightarrow  (k_{F}^{m} z)^{-1}\left[1+y/z
-2/(\pi k_{F}^{m} z) + {\cal O}(1/z^{2})\right]$. Inserting this in Eqs.(\ref{HF1}), 
the remaining integral can be evaluated analytically, yielding the 
important result   

\begin{equation}
V_{x}(z\rightarrow \infty )\rightarrow u_{x}^{m}(z\rightarrow \infty
)\rightarrow -\frac{e^{2}}{z}(1+\frac{\beta}{z}+...),
\label{HF2}
\end{equation}
with $\beta = \overline{z}\,^{m}-2/(\pi k_{F}^{m})$. It is interesting to note 
that no explicit knowledge of $\xi _{m}(z)$ is needed
in passing from Eq.(\ref{HF1}) to (\ref{HF2}), as just normalization 
has been used. Let us emphasize, however, that Eq.(\ref{HF2}) 
is an intrinsic slab result, as in its derivation the discrete character 
of the energy spectrum along the $z$ coordinate played a crucial role. This can 
be made more explicit by considering that $\overline{z}\,^{m} \sim d$ and 
$k_{F}^{m} \sim 1/d$, which allows
approximate the $\beta /z$ term in Eq.(\ref{HF2}) as proportional to $d/z$. For the 
expansion to be valid, this term should be smaller than the leading one, 
implying the slab limit $d/z < 1$.

\begin{figure}[h]
\includegraphics[width=\linewidth,clip=true,angle=0]{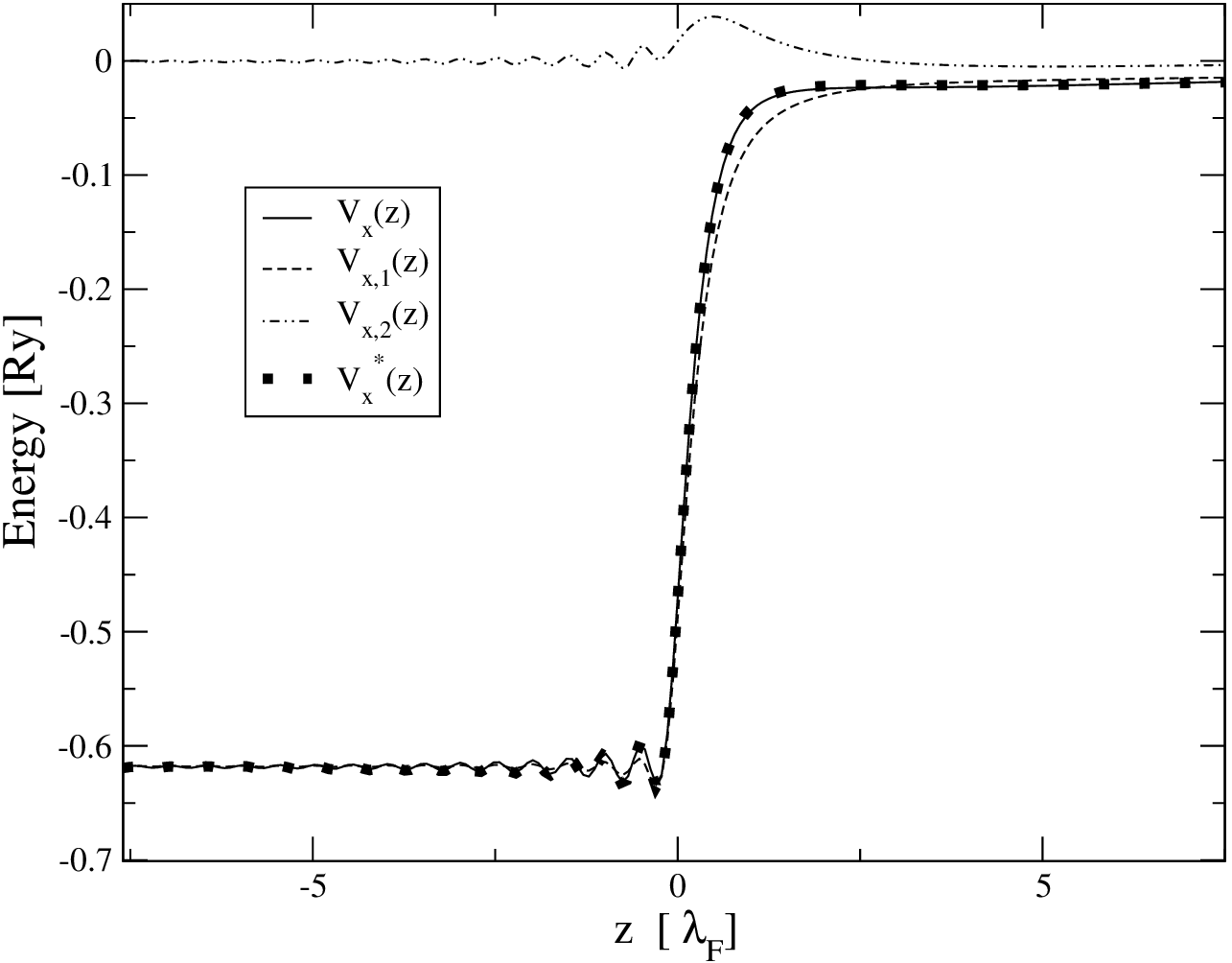}
\caption{ $V_{x}(z)$ in terms of the two components $V_{x,1}(z)$ 
and $V_{x,2}(z)$ for $r_{s}=2.07$ and 
$d=30 \lambda_{F}$. $V_{x}^{*}(z)$ (full dots) corresponds to the 
exact-exchange potential including correlation ({\it a la} LDA).} 
\end{figure}

\begin{figure}[h]
\includegraphics[width=\linewidth,clip=true,angle=0]{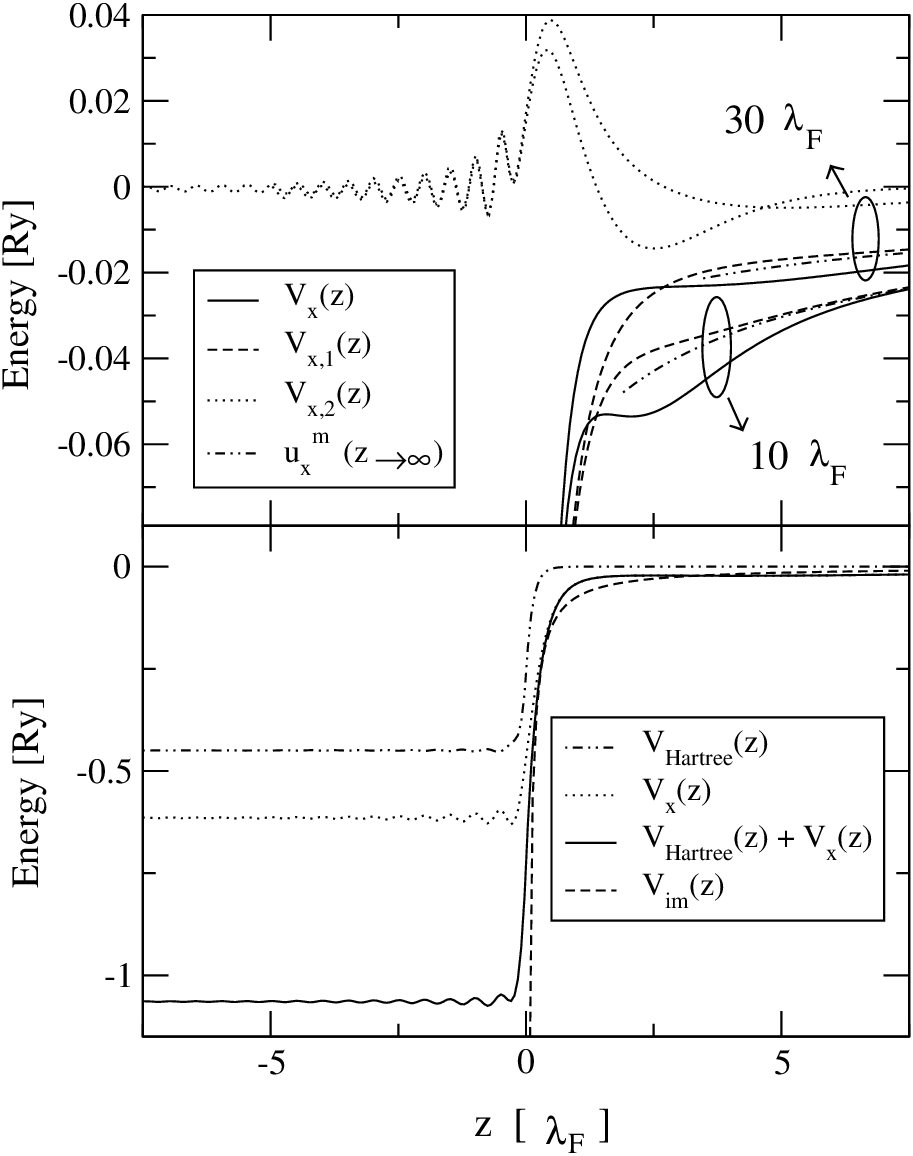}
\caption{Upper panel: details of $V_{z}(z),$ $V_{x,1}(z),$ $V_{x,2}(z),$ and 
$u_{x}^{m}(z\rightarrow \infty )$ in the vacuum region, for two slab sizes.
Lower panel: comparison of $V_{\text{KS}}(z)$ with $V_{im}(z)$, for $%
d=30\lambda _{F}.$ $r_{s}=2.07$ in both panels. }
\end{figure}

The top panel of Fig.3 displays the behavior of $V_{x}(z)$ in the
neighborhood of the metal-vacuum interface, for $r_{s}=2.07$, and two
different slab sizes. For the narrower slab, the asymptotic regime is
reached about $7$ $\lambda _{F}$'s from the jellium edge, resulting in an
excellent agreement between $V_{x}(z)$ calculated numerically, and the
asymptotic approximation of Eq.(\ref{HF1}). The oscillation which appears in
$V_{x}(z)$, is due to a crossover regime, where the density passes from being 
essentially dominated by $\xi _{m-1}^{2}(z)$ to $\xi _{m}^{2}(z)$.  
 For the slab with $d=30$ $
\lambda _{F}$, the asymptotic limit moves away from the jellium edge, and a
good matching is reached only between the KLI component $V_{x,1}(z)$ and $%
u_{x}^{m}(z\rightarrow \infty )$. However, and due essentially to the fact
that $V_{x,2}(z)$ has still a sizeable value for the largest $z$ coordinate
within the reach of the numerical calculation, not quantitative agreement is
observed yet between $V_{x}(z)$ and Eq.(\ref{HF1}). Having presented then
compelling numerical evidence of the validity of Eq.(\ref{HF1}) in
representing the exact-exchange potential in the asymptotic regime, the
result of Eq.(\ref{HF2}), which follows at once from Eq.(\ref{HF1}), 
is also confirmed numerically.

The long-standing puzzle of the image-potential is briefly discussed now at
the light of the results presented in the lower panel of Fig.3. Already in
1936, in his pioneering study of the surface barrier at the jellium
metal-vacuum interface, Bardeen considered that under the {\it combined} effects
of exchange and correlation, electrons far enough from the jellium edge
should be subject to the classical image-potential 
$V_{im}(z)=-e^{2}/4z$\cite{bardeen}. In fact, he imposed this asymptotic behavior on his
approximate Hartree-Fock calculation. Many years later, the first
application of DFT at the study of the same problem was performed by Lang
and Kohn\cite{lang}. They used LDA, so their $V_{xc}(z)$ vanishes
exponentially as $z\rightarrow \infty $, as already discussed. However, they
recognized that the correct $V_{xc}(z)$ would behave like the classical
image potential, 
\begin{equation} V_{xc}(z\rightarrow \infty )\rightarrow 
V_{im}(z)=-\,e^{2}/4z.  \label{image} 
\end{equation}
Motivated by this, we have included $V_{im}(z)$ in the lower panel of Fig.3,
and compared with our exact-exchange results. As expected, $V_{im}(z)$
decays more rapidly that our $V_{x}(z),$ and misses the shoulder which is
present in the exact-exchange solution. {\it Assuming} that Eq.(\ref{image}) is
correct, we speculate that the apparent discrepancy with Eq.(\ref{HF2})
is due to correlation, which is the only missing ingredient in our 
exact $x$-only calculation. This would imply that $V_{c}(z\rightarrow \infty)
\rightarrow 3e^{2}/4z$. We emphasize, however, that this conclusion 
is a direct consequence of the assumption that $V_{xc}(z\rightarrow \infty )
\rightarrow V_{im}(z)$. To the best of our knowledge, no rigorous proof inside the 
DFT framework exist for this equivalence\cite{dobson}. As a by-product, our 
contribution clearly points that more work is needed on this subtle issue.

Let us place our results in the context of other related works. In Ref.\cite
{ulf} the asymptotic behavior of $V_{xc}(z)$ for the case of a semi-infinite
jellium surface was addressed from a many-body point of view. It was stated
(without proof), that for macroscopic systems the exchange potential tends
exponentially to zero, concluding that the long-range components of $%
V_{xc}(z)$ can only originate from correlation effects. Similar conclusions
were reached in Ref.\cite{sham} by analyzing the asymptotic limit of the
Sham-Schl\"{u}ter integral equation for $V_{xc}(z)$, except for the result
that $V_{x}(z\rightarrow \infty )\rightarrow -1/z^{2}$, in disagreement with
Ref.\cite{ulf}. Being our numerical calculations restricted to finite slab
geometries, we can not compare with these two ones. We would like to
address, however, that our result that $V_{x}(z\rightarrow \infty
)\rightarrow -\,e^{2}/z$ for very thick slabs (in terms of $\lambda _{F}$),
is not in agreement with either two results. The slab calculations of Ref.\cite
{krotscheck} and Ref.\cite{eguiluz}, are much closer to ours. In Fig.10 of
Ref.\cite{krotscheck} results are shown for $V_{x}(z)$, obtained through the
numerical solution of the OEP integral equation, for a slab of $4\,\lambda
_{F}$ width ($r_{s}=2.07$). The figure suggests that $V_{x}(z)=0$ was forced
about $1\,\lambda _{F}$ from the jellium edge, spoiling any detailed study
of the asymptotic properties of $V_{x}(z).$ Using the same approach, Fig.2
of Ref.\cite{eguiluz} present results for $V_{x}(z)$ for slab thickness of
about $5\,\lambda _{F},$ for $r_{s}=3.23.$ From the asymptotic analysis of
the numerical results in their vacuum region, that only extends 1.25 $%
\lambda _{F}$ from the jellium edge, the authors of Ref.\cite{eguiluz}
conclude that $V_{x}(z\rightarrow \infty )\rightarrow -1/z^{2}.$ The results
presented above suggest, however, that the correct asymptotic behavior sets
in at much larger distances from the jellium edge. Also, the shoulder is not
discernible in their results for $V_{x}(z).$ Our work is also not in
agreement with the results of Ref.\cite{solomatin}, where through
approximate analytical techniques applied to a model slab geometry, it is
claimed that $V_{x}(z\rightarrow \infty )\rightarrow -1/z^{2}$ asymptotically.

What about correlation? Both limits $V_{x}(-d/2)\rightarrow V_{x}($3D$)$
and $V_{x}(z\rightarrow \infty )$ $\rightarrow -e^{2}/z$ are unchanged if
correlation is included. The rigidity of the bulk-like limit is displayed in
Fig.2, where it is seen that $V_{x}(z)$ does not change in the metallic region
 if correlation is present. 
 This is essentially a consequence of the boundary condition $
\overline{V}_{x}^{m}=\overline{u}_{x}^{m},$ that ensures the exact
fulfillment of the exchange bulk-like limit, independently of correlation
effects. The asymptotic result of Eqs.(\ref{HF1}) and (\ref{HF2}) are also
rigorously valid even if correlation is included, as in this case 
each one of the basic Eqs.(\ref{xc1})-(\ref{shift}) can be 
splitted in an exchange and correlation components, due to the fact that $E_{xc}=
E_{x}+E_{c}$. All the subsequent derivations leading to Eq.(\ref{HF2}) for the 
exchange component of the total KS potential remains valid in consequence in 
the presence of correlation, that will not modify the general properties of the
$\xi_{i}(z)^{'}$s and the exchange component of the $\psi _{i}(z)^{'}$s
  on which they are based, such as asymptotic behavior and
 normalization.

In summary, we have achieved a rigorous analytical and numerical study of
the exchange component of the surface barrier at the jellium metal-vacuum
interface. The Kohn-Sham exact-exchange potential develops a shoulder-like
structure within $1-2\;\lambda _{F}$'s from the jellium edge, and decays as $%
-\,e^{2}/z$ at much larger distances. This exchange asymptotic behavior is
unperturbed by correlation. With these exact results at the exchange level, 
the challenge quest of DFT for a compatible energy correlation functional 
is now much better focalized.

CH and SR acknowledge financial support from CONICET and CNEA/CONICET, 
respectively. This work was partially supported by the CONICET under grant
PIP 2753/00 and the ANPCyT under PICT02 03-12742.


\begin{references}
\bibitem{units}  Numerical results are presented using the atomic Rydberg
and $\lambda _{F}=2\pi /k_{F}$ as units of energy and length, respectively.
Jellium densities are defined through the dimensionless parameter $%
r_{s}=(3/4\pi \overline{n}a_{0}^{3})^{1/3}.$ $a_{0}$ is the Bohr radius.
\end{references}
\end{document}